# Dilute Rhenium Doping and its Impact on Intrinsic Defects in MoS$_2$

Riccardo Torsi,[1] Kyle T. Munson,[2] Rahul Pendurthi,[3] Esteban A. Marques,[4,5] Benoit Van Troeye,[4] Lysander Huberich,[6] Bruno Schuler,[6] Maxwell A. Feidler,[1] Ke Wang,[7] Geoffrey Pourtois,[4] Saptarshi Das,[3] John B. Asbury,[2,†] Yu-Chuan Lin,[1,#] Joshua A. Robinson[1,2,7,8*]

1. Department of Materials Science and Engineering, The Pennsylvania State University, University Park, Pennsylvania 16802, United States
2. Department of Chemistry, The Pennsylvania State University, University Park, Pennsylvania 16802, United States
3. Department of Engineering Science and Mechanics, The Pennsylvania State University, University Park, Pennsylvania 16802, United States
4. Imec, Leuven 3001, Belgium
5. Department of Molecular Design and Synthesis, KU Leuven, Celestijnenlaan 200f - Postbox 2404, 3001 Leuven, Belgium
6. nanotech@surfaces Laboratory, Empa-Swiss Federal Laboratories for Materials Science and Technology, Dübendorf 8600, Switzerland
7. Materials Research Institute, The Pennsylvania State University, University Park, PA, 16802, USA
8. Department of Physics, The Pennsylvania State University, University Park, Pennsylvania 16802, United States

\# yul194@psu.edu

† jba11@psu.edu

\* jar403@psu.edu

## Abstract

Substitutionally-doped 2D transition metal dichalcogenides are primed for next-generation device applications such as field effect transistors (FET), sensors, and optoelectronic circuits. In this work, we demonstrate substitutional Rhenium (Re) doping of MoS$_2$ monolayers with controllable concentrations down to 500 parts-per-million (ppm) by metal-organic chemical vapor deposition (MOCVD). Surprisingly, we discover that even trace amounts of Re lead to a reduction in sulfur site defect density by 5-10×. *Ab initio* models indicate the free-energy of sulfur-vacancy formation is increased along the MoS$_2$ growth-front when Re is introduced, resulting in an improved stoichiometry. Remarkably, defect photoluminescence (PL) commonly seen in as-grown MOCVD MoS$_2$ is suppressed by 6× at 0.05 atomic percent (at.%) Re and completely quenched with 1 at.% Re. Furthermore, Re-MoS$_2$ transistors exhibit up to 8× higher drain current and enhanced mobility compared to undoped MoS$_2$ because of the improved material quality. This work provides important insights on how dopants affect 2D semiconductor growth dynamics, which can lead to improved crystal quality and device performance.

## Introduction

The development of conventional semiconductor technology was spurred by the ability to modulate the transport properties of electronic materials through replacement of atoms in the lattice of a host material with foreign atoms (referred to as doping). Complementary metal-oxide semiconductor (CMOS) integrated circuits rely on n-type and p-type doping of Si through the controlled substitution of Si atoms with P and B, respectively. Semiconducting transition metal dichalcogenides (TMDCs) are the 2D analog to traditional semiconductors, such as Si and GaAs, making them of interest for photon sensors, atomically-thin quantum wells, and transparent, flexible electronics.[1–4] While there is progress in the realization of wafer scale epitaxial 2D TMDCs,[5–9] the material selection for dielectric overlayers[10] and substrates[11] can have tremendous effects on carrier concentrations and resulting electrical conductivity. Moreover, a lack of scalable defect and doping control methods impede the implementation in many application spaces.[12–14]

Substitutional doping is emerging as the most viable route for stable incorporation of foreign atoms into a 2D host's metal or chalcogen site to tune carrier concentration or impart novel functionalities in TMDCs.[13,14] To-date, most *in situ* metal-site substitutional doping is achieved by thermally vaporizing powders in a chamber or pyrolyzing liquid phase precursors spin-coated to substrates that provide dopants during growth.[15] These methods can realize p-type[16] and n-type[17,18] substitutional doping, but powder-based CVD is known to exhibit poor film uniformity and dopant concentration variability.[19] MOCVD, on the other hand, achieves p-type[20] and n-type[21] *in situ* doping of TMDCs with wafer-scale uniformity[7,8,22] thereby matching industrial quality standards. In general, however, the ionization energies of dopants in 2D materials are on the order of a few hundred meV due to quantum confinement and reduced charge screening effects[23,24] - an order of magnitude higher than bulk semiconductors. As a result, dopant concentrations of >1 at.% - well beyond that of traditional doping - are required to modify the transport properties of 2D TMDCs[25,26] thereby limiting our understanding of optoelectronic properties of TMDCs in the dilute dopant regime. In this vein, Re is an ideal transition metal dopant to study as it forms shallower donor levels[27,28] relative to other n-type dopants (e.g. Mn, Fe)[13] in 2D TMDCs.

In this paper we demonstrate that dilute dopant concentrations can have dramatic effects on vacancy formation and impurities at S sites of MoS$_2$ monolayers. First, we establish the growth of uniform Re-doped MoS$_2$ monolayers on sapphire substrates using Mo(CO)$_6$, H$_2$S, and Re$_2$(CO)$_{10}$. We use a combination of X-ray photoelectron spectroscopy (XPS), Z-contrast scanning transmission electron microscopy (Z-STEM), and laser ablation inductively coupled plasma mass spectrometry (LA-ICPMS) to confirm that Re-concentration is controlled down to 500 ppm by tuning Re$_2$(CO)$_{10}$ during growth. Z-STEM experiments elucidate that Re dopants only substitute at the Mo sites (Re$_{Mo}$) and help reduce the defect density at S sites in MoS$_2$. Low-temperature scanning tunneling spectroscopy (STS) confirms the n-type character of Re$_{Mo}$. We find that even small Re- doping levels ($\leq 0.1$ at.%) suppress defect-related emission in both time-resolved and temperature-dependent photoluminescence (PL) spectroscopy. The reduction of intrinsic defects agrees well with our *ab initio* model of the MoS$_2$ growth front. When Re is present at the growth front of a MoS$_2$ grain, the energy of formation of edge sulfur vacancies is increased by ≈50 kJ/mol. Finally, back gated field-effect transistors (BGFET) reveal that defect suppression induced by Re incorporation also results in improved transport properties in the low doping regime.

## Results and Discussion

Monolayer (ML) MoS$_2$ films with controlled Re incorporation are synthesized on c-plane sapphire at 1000 °C in a custom-built MOCVD reactor (**Fig. 1a**). A multi-step growth process[8] (**Fig. S1**) with separate nucleation, ripening, and lateral growth stages is used to regulate the nucleation rate (see Methods section). Controlled Re incorporation in the MoS$_2$ films is achieved by introducing Re$_2$(CO)$_{10}$ in the growth chamber during MOCVD. Re$_2$(CO)$_{10}$ is chosen as the dopant precursor due to its suitable vapor pressure[29] and clean pyrolysis.[30] A coalesced, uniform 1cm$^2$ ML film (**Fig. 1a, inset**) with a low percentage of bilayer islands (**Fig. 1b**) is deposited in approximately 30 minutes, where the underlying morphology in the monolayer regions (**Fig. 1b**) originates from the substrate sapphire step edges. The Re concentration in doped MoS$_2$ samples is confirmed by XPS quantification of Re 4$f_{7/2}$ and Re 4$f_{5/2}$ peaks (**Fig. S2**). By modulating H$_2$ carrier gas flow through the Re$_2$(CO)$_{10}$ bubbler, Re content in MoS$_2$ is tuned from alloying (> 6 at.%) to doping levels (< 0.1 at.%). Due to resolution limitation of the XPS instrument, laser-ablation inductively coupled plasma mass spectrometer (LA-ICPMS) was used to detect lower Re content. The concentration curve in **Fig. 1c** demonstrates agreement between various compositional analysis techniques and a linear relationship between the $Re_2(CO)_{10}$ flow and Re content in the $MoS_2$ samples. Given the linearity of the concentration curve, extrapolation can be used to realize films with ppm-level dopant concentrations. High-angle annular dark field (HAADF) Z-STEM (**Fig. 1d** and inset, and **Fig. S3**) confirms that Re substitutes Mo, and is uniformly distributed in $MoS_2$ at the concentrations examined here. The Re atoms present in a representative STEM image (20 x 20 $nm^2$) corresponds to a Re concentration of 1.1 at.%, agreeing well with fitting the XPS Re 4$f$ peaks. Furthermore, the valence band maximum (VBM) measured by XPS evolves with Re concentration (**Fig. 1e**), indicating the Fermi level of $MoS_2$ is impacted by Re incorporation. The VBM position gradually shifts toward higher binding energy values with increasing Re concentrations from 1.24 eV to 1.55 eV in the pristine $MoS_2$ and 1 at.% Re-$MoS_2$ films, respectively, indicating Re is n-doping $MoS_2$. Low-temperature (5 K) scanning tunneling spectroscopy (STS) confirms Re introduces electronic states near the conduction band of $MoS_2$. Furthermore, evaluating the electronic structure of single Re substituting for Mo grown on quasi-freestanding epitaxial graphene on SiC substrates (Re-$MoS_2$/QFEG/SiC) using STS reveals that the Re is positively charged (ionized) $Re_{Mo}^+$ (**Fig. S4a**, inset). Differential conductance (d$I$/d$V$) spectra recorded on a single $Re_{Mo}^+$ (blue curve in **Fig. S4a-b**) demonstrate that Re impurities introduce several defect resonances below the conduction band minimum labelled A-E in **Fig. S4b**. Notably, the lowest unoccupied defect state is close to the Fermi level (at 0 V), which leads to tip-induced charging peaks at low negative bias voltages.[31] The spatial distribution of the main defect states shown in **Fig. S4c** are similar to the $Re_W^+$ states in previously reported $WSe_2$[21].

Incorporation of Re impacts the structure and absorption properties of the host $MoS_2$ material.[20,21] We established the Re dopant concentrations-structure relationship in $MoS_2$ with Raman spectroscopy and optical absorption measurements. Under 633 nm excitation, pristine ML $MoS_2$ exhibits standard Raman peaks for in-plane ($E'$ at 386 $cm^{-1}$) and out-of-plane ($A_1'$) at around 405 $cm^{-1}$) modes and also low-intensity features in the 160-270 $cm^{-1}$ spectral range related to either defects[32,33] or dopants[20,21] (**Fig. 1f**). As Re concentration increases from pristine to 1 at.%, the second order bands of $MoS_2$[32] including 178 (TA), 189 (ZA), 230 (LA), 417, and 460 $cm^{-1}$ (2LA) increase accordingly. Nevertheless, small Re concentrations will not significantly alter the phonon properties compared to pristine $MoS_2$. This observation is consistent with the previous results of substitutionally-doped $WSe_2$.[20,21]. The optical absorption spectrum of both pristine and Re-$MoS_2$ exhibits the characteristic A, B, and C excitons in the visible wavelength range (**Fig. S5**). While these exciton positions change negligibly with increasing Re concentration, both A and B bands become broader and weaker from 1 to 10 at.% Re due to increasing fraction of Re-S bonding in the MoS$_2$ lattice and Re-related energy states.[34] Since ML ReS$_2$ has an indirect bandgap at 1.65 eV,[35] the transition from MoS$_2$ to an Mo$_{1-x}$Re$_x$S$_2$ alloy will shift absorption peaks toward longer wavelengths (smaller bandgaps) with increasing Re content. To avoid excessive change in the structure and properties of MoS$_2$, Re concentration for samples in this work is kept ≤ 1 at.%.

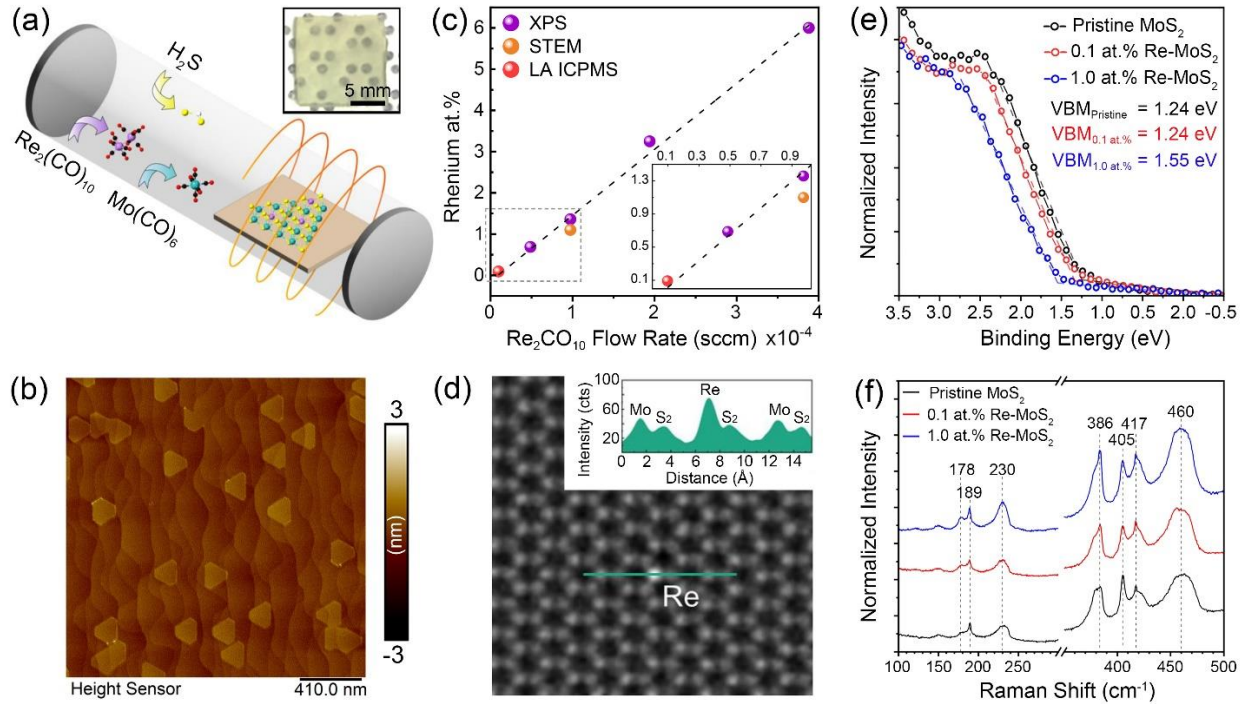

**Figure 1. Synthesis and properties of ML Re-MoS$_2$:** (a) Schematic illustration of MOCVD process for ML Re-MoS$_2$ films deposited on c-plane sapphire. Inset: camera picture of a Re-MoS$_2$ film on 1 cm$^2$ sapphire. (b) AFM topography of ML Re-MoS$_2$. (c) The plot of Re$_2$(CO)$_{10}$ *versus* Re at.% verified by a variety of surface characterization techniques. Inset: A close view of the plot highlights the smaller amounts of Re in MoS$_2$ grown with the corresponding Re$_2$(CO)$_{10}$ flow. (d) Atomic resolution STEM image shows a Re substituting Mo in MoS$_2$. Inset: Line profile highlighting the relative Z-intensity of Re, Mo, and S2. (e) VBM edges of pristine and Re-MoS$_2$ at 0.1 and 1 at.% measured by XPS. (f) Raman spectral evolution of ML MoS$_2$ with 0, 0.1, and 1 at.% Re under 633 nm excitation.

Our Z-STEM image analysis demonstrates that Re-doping reduces the S-site defects in MoS$_2$ films. We counted vacancies and impurities in the sulfur lattice on seven 100 nm$^2$ STEM images from both pristine and Re-MoS$_2$ films (**Fig. 2a-b** and **Fig. S6**). Based on their Z-intensity, we confirmed the presence of un-passivated sulfur vacancies (V$_S$), V$_S$ passivated by carbohydrate (CH$_S$) and oxygen (O$_S$)[36], and double-S vacancies (V$_{2S}$) in pristine MoS$_2$ (**Fig. S7**). V$_S$ passivation with CH or O is kinetically favorable[37] and might occur either inside the growth chamber or in the atmosphere.[38] Statistical analysis of the STEM images indicates a dramatic reduction in S-site defects from $\approx 2.7 \cdot 10^{13}$ cm$^{-2}$ to $5.1 \cdot 10^{12}$ cm$^{-2}$ in pristine MoS$_2$ and Re-MoS$_2$, respectively (**Fig. 2c**). While chalcogen vacancy suppression in TMDCs by doping (Re-WS$_2$ crystals[39]) and alloying (2D Mo$_{0.82}$W$_{0.18}$Se$_2$[40]) is known, there's no clear explanation for the driving force of such suppression during material synthesis. We also note that, different from previous reports, unintentional defect reduction even occurs in samples $\leq 0.1$ at.% based on the results from our optical characterizations that will be discussed below.

Density functional theory (DFT) modeling of the MoS$_2$ growth front reveals a significant increase in formation energy of sulfur vacancies during synthesis when Re is incorporated at the grain edge. Z-STEM images of Re-MoS$_2$ nuclei grown on a graphene TEM grid (**Fig. 2d**) indicate Re atoms bind to the domain edge with a small number of Re dopant atoms already incorporated within the nuclei. This observation is consistent with *ab-initio* thermodynamic calculations that indicate Re atoms have a preferential edge attachment during growth (**Fig. S8**). Considering this attachment behavior, we theoretically investigate the formation energy for a V$_S$ located at the edge of the crystal with varying distance with respect to a Re dopant (**Fig. 2e,** inset), and find that V$_S$ formation energy increases when the Re is getting closer to the V$_S$ (**Fig. 2e**). When Re is closest to the edge, the formation energy is the highest around 210 kJ mol$^{-1}$. On the other hand, the formation energy begins to decrease as the Re moves away from the edge and converges to $\approx 163$ kJ mol$^{-1}$. Therefore, it is more energetically expensive to introduce a vacancy into MoS$_2$ at the edge when Re is present. The relative defect concentration of V$_S$ can be estimated via:

$$\frac{n_{edge,Re-V_S}}{n_{edge,V_S}} = \exp\left(-\left(\frac{E_{form,\,Re\,edge}-E_{form,edge}}{RT}\right)\right),$$

where $E_{form,Re\,edge}$ and $E_{form,edge}$ are the V$_S$ formation energy at the edge with Re in its close vicinity or far away, respectively, while $n_{edge,Re-V_S}$ and $n_{edge,V_S}$ are their corresponding concentrations, R denotes the molar gas constant, and T is growth temperature. Taking the growth temperature at 1273 K as example, and assuming the last point of the formation energy in **Fig. 2e** is representative when Re is far away, the reduction of defects is estimated to be 98 %. Therefore, we hypothesize that V$_S$ are incorporated into MoS$_2$ during its lateral growth, and this process is hindered by the incorporation of Re at the MoS$_2$ edge. This result is quite remarkable as it contradicts previous theory work on the Re-MoS$_2$ system which indicates that V$_S$ in the MoS$_2$ basal plane have a lower formation energy next to Re-dopants resulting in stable Re-$V_S$ complexes.[28,41] It is worth noting that such complexes are rarely observed in our Z-STEM images suggesting that the growth-front effects dominate over the formation of $V_S$ in the basal plane. This discrepancy emphasizes the importance of considering how dopant atoms are being incorporated during material synthesis and how these affect defect formation mechanisms.

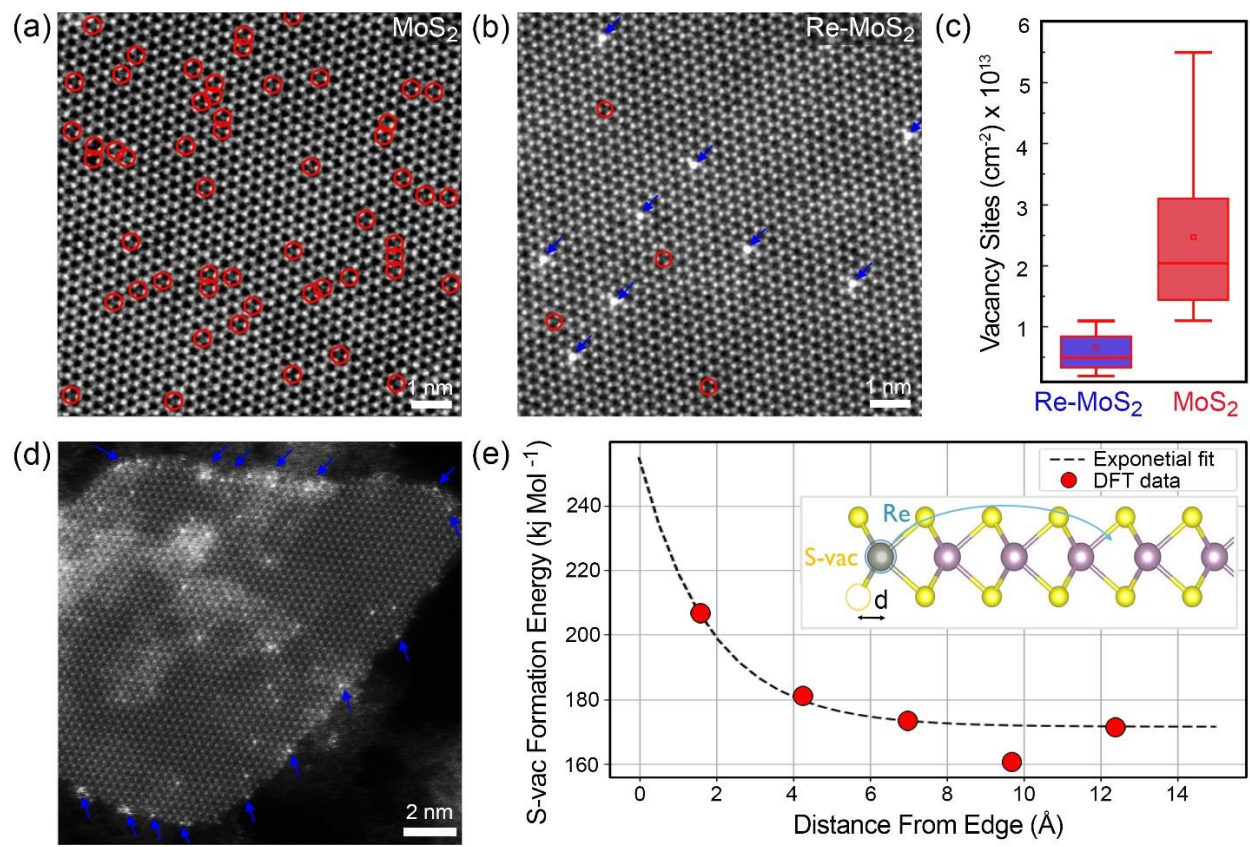

Figure 2. TEM analysis of sulfur vacancies in ML Re-MoS$_2$: (a,b) the point defect density analysis based on Z-contrast STEM images of (a) MoS$_2$, and (b) Re-MoS$_2$ indicates their S-site defect (marked with red circles) densities are (c) approximately $3 \cdot 10^{13}$ cm$^{-2}$ and $5 \cdot 10^{12}$ cm$^{-2}$, respectively. (d) Z-contrast STEM image of a ML Re-MoS$_2$ domain at the initial growth stage reveals a high density of Re atoms and Re clusters (marked with blue arrows) attached to the domain edge. (e) DFT model of $V_S$ formation energy as a function of the distance between the $V_S$ at the edge of MoS$_2$ model and the Re position moving away from the edge, and the corresponding energy values as a function of the Re position. In the box plots (c) the lower and upper ranges represent the lower and upper quartile, respectively while the whiskers are the outliers. Within each box, the horizontal line is the median value and the data point is the average.

Photoluminescence (PL) is sensitive to defects and Re doping even in the dilute limit (**Fig. 3a**). The room temperature (RT) PL spectra of pristine (0), 0.05, and 0.1 at.% Re-MoS$_2$ are fit with two Lorentzian curves each, centered at 1.86 and 1.90 eV. The PL spectra of 1 at.% Re-MoS$_2$ is fit with curves at 1.79 and 1.83 eV, respectively. The ≈ 40 meV energy separation between the fitting curves within the main PL spectra is consistent with that of trions and neutral A excitons.[42] Using a mass action model,[43] the electron density in the films is estimated from the intensity ratio of the trion and A-exciton PL curves (**Fig. 3b**), and is found to increase from $5.0 \cdot 10^{12}$ to $3.3 \cdot 10^{13}$ cm$^{-2}$ in a square-root-like trend from pristine to 1 at.% Re-MoS$_2$. The error bars represent the uncertainty limits of the electron densities obtained from the fitting procedure. The increase in electron concentration is consistent with the VBM shift measured in XPS and the n-type character of Re-dopants in STS. The increased electron density is accompanied by a modification in exciton recombination kinetics. Planar frequency-time maps of time-resolved photoluminescence (TRPL) spectra measured at RT under identical excitation conditions (445 nm, 5 pJ/pulse) are shown in **Fig. 3c**. The decay traces of the TRPL map further highlight the differences between the recombination kinetics (**Fig. 3d**), where each exhibit an initial, fast component (0–500 ps) and a slow component (500–5000 ps). DFT calculations and TRPL of MoS$_2$ suggest the radiative lifetime of excitons is 270 - 470 ps.[44–46] Therefore, we attribute the initial fast component to the radiative recombination of free excitons (Inset, **Fig. 3d**). Compared to pristine MoS$_2$, the faster decay of free excitons in Re-MoS$_2$ arises from accelerated non-radiative recombination via excess carriers (trions).[47] Exciton lifetime enhancement in pristine MoS$_2$ beyond ≈ 500 ps arises from unintentional defects commonly found in both exfoliated and CVD-grown MoS$_2$ films.[44,47] For example, measurements of CVD-grown MoS$_2$ revealed enhanced PL lifetimes at misaligned grain boundaries.[47] Here, biexponential functions are used to fit the fast (free excitons) and slow (defect-related) components of the TRPL decay traces (**Fig. 3d**) to quantify their average decay times (black lines, **Fig. 3d**). The best-fit parameters of the functions appear in **Table S1**, from which we obtained average exciton recombination lifetimes of 307, 121, and 91 ns for 0, 0.1, and 1 at.% Re-MoS$_2$, respectively. We note there is significant reduction in the defect-related component of the Re-MoS$_2$ TRPL decays (**Table S1**). The slow component reduction may be due to either increased trap filling as a result of the raised Fermi level by Re doping, thereby reducing the concentration of gap states available to assist in the slow recombination processes,[48] or due to a lower density of unintentional point defects as Re is incorporated, as demonstrated in our previous defect analysis.

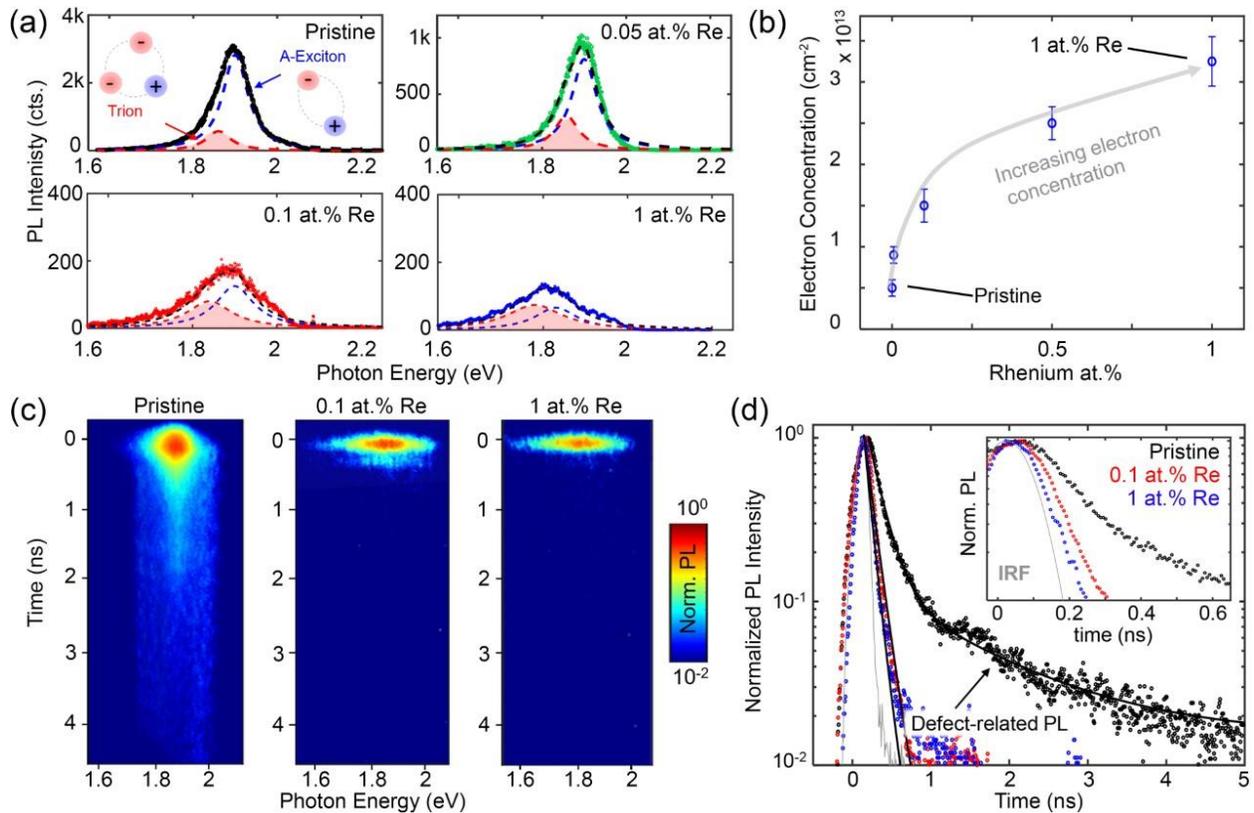

**Figure 3. Room-temperature optical characterizations for Re-MoS$_2$:** (a) Photoluminescence (PL) spectra of pristine and Re-doped MoS$_2$ films. The PL spectra were fit with two Lorentzian curves to determine the contribution of A-excitons (blue dashed line) and trions (red dashed line) to the PL spectra. (b) Electron density calculated by a mass action model (see **Fig. S9**) plotted as a function of Re at.%. (c) Two-dimensional frequency-time plots of 0, 0.1, and 1 at.% Re films measured following optical excitation at 445 nm. (d) Exciton recombination kinetics obtained by integrating the PL spectra between 1.7 eV and 1.9 eV. The comparison demonstrates that excitons in Re-MoS$_2$ films undergo faster recombination.

Temperature-dependent PL measurements provide direct evidence for quenching of defect-related emission in Re-MoS$_2$ and show that the phenomenon is persistent even in dilute-doped films. Low-temperature measurements are critical for resolving defect-related luminescence as trapped exciton emission begins to dominate over the band-edge contribution at < 77 K.[49] Unintentional defects can have a broad spectrum of effects on the PL of 2D TMDCs. Chalcogen vacancies, whether they be un-passivated or filled, can act as active sites for molecular adsorption of gas molecules such as O$_2$, H$_2$O, or N$_2$.[50] This molecular adsorption mechanisms gives rise to sub-gap luminescence sometimes referred to as L-band.[50] Furthermore, chalcogen vacancies can introduce mid-gap defects levels in the band-gap which effectively trap excitons and also result in sub-gap luminescence upon radiative recombination. Here, we do not specify whether our measured sub-gap emission originates from molecular adsorption at defects or bound states because both lead to similar optical effects, instead we broadly refer the sub-gap emission as "defect-related emission ($X_D$)". **Fig. 4a** displays PL offset spectra of a MoS$_2$ film collected from 77 to 300 K following optical excitation at 445 nm. At ≤ 200 K, the spectra exhibit emission from both free excitons (≈1.9 eV) and defect-related emission (≈1.65 eV).[44,51,52] At 77 K, the broad $X_D$ dominates, indicating a distribution of defects and impurities[53–56] in MoS$_2$ including V$_S$, V$_{2S}$, and passivated V$_S$. This result is consistent with the literature[56,57] and our projected density of states (PDOS) calculations **(Fig. 4d)** which show that V$_S$ introduce defect energy states inside the bandgap of MoS$_2$. The temperature-dependent PL **(Fig. 4b-4c)** of 0.05 and 1 at.% Re-MoS$_2$ films exhibits significantly reduced $X_D$ emission for 0.05 at.% Re-MoS$_2$ and it is completely quenched at 1 at.% Re at 77 K. Additionally, the PL spectra of free excitons in the 1 at.% Re-MoS$_2$ film is shifted by ≈0.1 eV compared to the pristine analog. We examined the PDOS of Re-MoS$_2$ to understand the coupling between Re and V$_S$ and their impact on MoS$_2$ PL **(Fig. S10)**, which indicates that neutral Re atoms provide shallow donor states below the CBM of MoS$_2$.[18,21,28] These states effectively contribute to ≈ 100 meV reduction in bandgap size observed in our PL spectra for 1 at.% Re-MoS$_2$ **(Fig. 4c)**. **Fig. 4e** displays the PDOS of Re-MoS$_2$ with one V$_S$ as a function of Re-V$_S$ distance. When Re atoms neighbor V$_S$ (≈ 2.4 Å), deep mid-gap states are observed in the PDOS.[28,58] However, Re$_{Mo}$-V$_S$ complexes[58] are rarely observed in our TEM **(Fig. S6)**. As the Re-V$_S$ distance is increased to ≥ 3.9 Å, the localized energy states get closer to the CBM and eventually merge with Re donor states. Therefore, the quenching of $X_D$ in Re-MoS$_2$ is likely a convoluted combination of S-site defect density reduction, increased Fermi level energy above the defect level[59], and defect-state modulation by Re dopants.

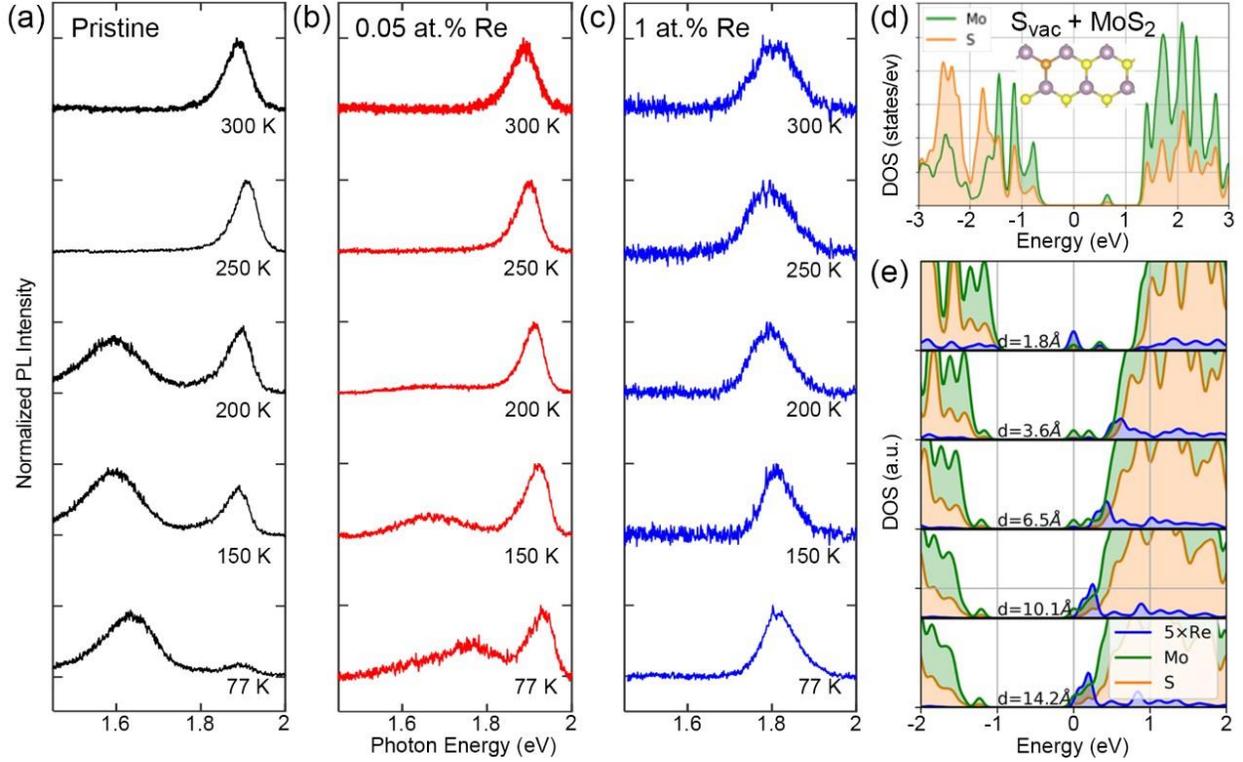

**Figure 4. Understanding the defect-related PL emission in Re-MoS$_2$:** (a) Temperature-dependent PL spectra from a) pristine MoS$_2$, b) 0.05 at.% Re-MoS$_2$, and c) 1.0 at.% Re-MoS$_2$. The low-temperature PL spectrum of pristine MoS$_2$ exhibits a broad PL emission band at 1.65 eV that arises from defect-related excitons ($X_D$), which is quenched significantly with increasing Re doping. (d) PDOS for perfect Re-MoS$_2$. (e) PDOS for Re-MoS$_2$ with one V$_S$ as a function of Re-V$_S$ distance. The Fermi level is aligned to 0 and the Re projection is amplified for sake of legibility.

Doping MoS$_2$ with Re yields improvement in transistor performance. This is evident when comparing BGFETs of MoS$_2$ with and without Re (see transport characteristics in **Fig. 5**). While the non-linearity of all drain current ($I_{DS}$) *versus* drain voltage ($V_{DS}$) curves strongly indicates the device performance is contact limited (Schottky barrier at the metal-film interface), the 0.1 at.% Re-MoS$_2$ BGFET exhibits a maximum $I_{DS}$ of ≈ 61.5 µA/µm, nearly 10× higher than pristine MoS$_2$ (≈ 7.6 µA/µm), which may be attributed to the reduction in MoS$_2$ lattice defects when Re is introduced to MoS$_2$. Moreover, the field effect mobility ($\mu_e$) is moderately increased from 4.1 cm$^2$/Vs in the pristine BGFETs to 8.8 cm$^2$/Vs in the 0.1 at.% Re-MoS$_2$ devices. However, when the Re content is further increased to 1.0 at.%, the maximum $I_{DS}$ and $\mu_{FET}$ reduces by nearly 30% due to increasing impurity scattering with more Re in the device channel. Furthermore, we find the $I_{DS}$ *versus* back-gate voltage ($V_{BG}$), at different $V_{DS}$ display dominant n-branch (device is in the ON state at $V_{BG} > V_{TH}$) across all samples, with improved uniformity with dilute Re doping

(**Table S2**). A box plot from the extracted ON-current ($I_{ON}$) values (**Fig. S11**) demonstrates that the $I_{ON}$ and $\mu_e$ enhancement with Re doping is statistically significant, which can be attributed to both n-type doping and to the reduction of S-site defects[56] in ML $MoS_2$.

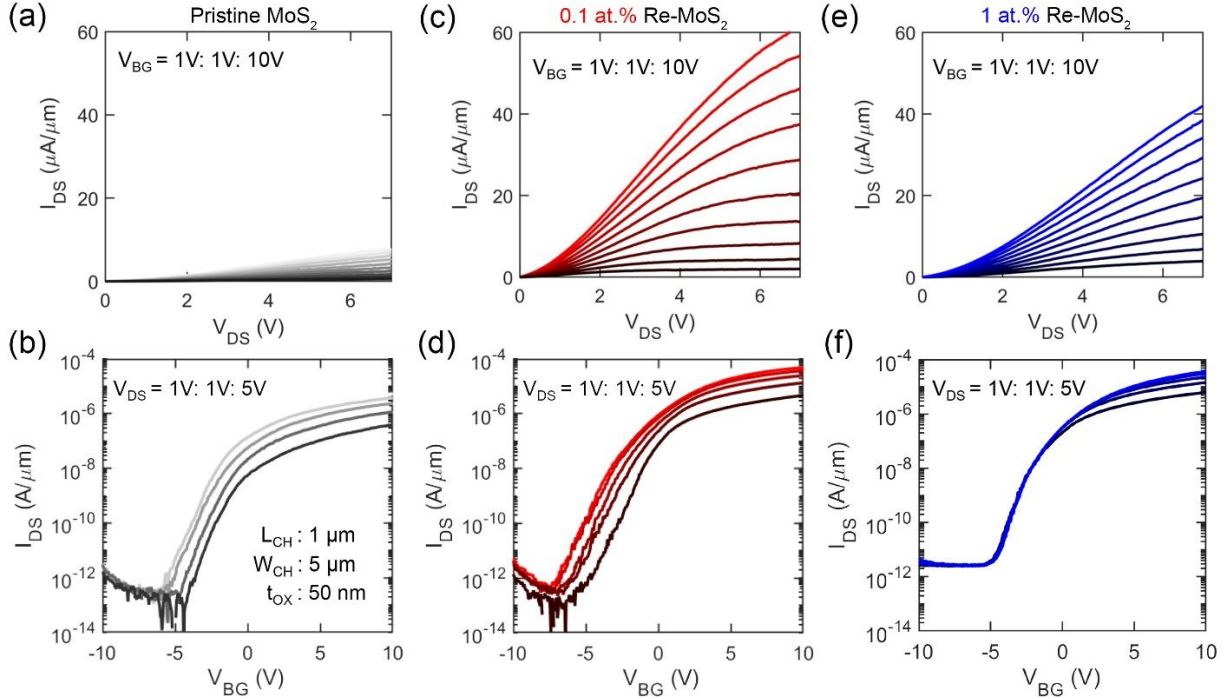

**Figure 5. Transport characteristics of pristine and Re- MoS₂ BGFETs:** Output characteristics ($I_{DS}$ *versus* $V_{DS}$) at different $V_{BG}$ values and transfer characteristics ($I_{DS}$ *versus* $V_{BG}$) at different $V_{DS}$ values for 0 (pristine) (a,b), 0.1 (c,d), and 1.0 at.% Re-MoS₂ (e,f) BGFETs. The devices were fabricated on $Al_2O_3$/Pt/TiN/p++-Si substrates. The device dimensions are shown inside (b).

## Conclusion

We have demonstrated controlled Re doping of $MoS_2$ down to hundreds of ppm (< 0.1 at.%). We confirm that Re atoms substitute Mo atoms in ML $MoS_2$ films and are uniformly distributed across the sample when grown on sapphire. The n-type nature of $Re_{Mo}$ is validated by STS and DFT calculations. The $MoS_2$ edge model suggests that $V_S$ formation at domain edges is hindered by the presence of Re, leading to a ≈ 5x reduction in S-site defects in Re-MoS₂ compared to pristine $MoS_2$, as observed in Z-STEM. This defect reduction is supported by RT-TRPL and 77 K PL measurements highlighting that $X_D$ can be effectively suppressed and even completely quenched at 77 K with Re at 0.05 and 1 at.%, respectively. Although dopant ionization is large for monolayers,[21,23,26,60] a small number of substitutional dopants in 2D TMDCs can have an immense impact on defect formation and properties of their hosts. This is particularly evident in the transport measurements of Re-MoS$_2$ BGFETs which show three-fold and two-fold improvement on $I_{ON}$ and $\mu_e$, respectively, compared to the pristine ones. While Re doping can reduce the S-site defects, the doping concentration needs to be carefully tuned in order not to offset the performance boost with excessive dopants. This work demonstrates how dilute substitutional dopant concentrations can substantially reduce the unintentional defect concentration in 2D TMDCs alleviating their detrimental effects.

## Experimental Methods

**MOCVD Growth and Doping:** MoS$_2$ films are grown in a custom-built horizontal, hot-wall metal organic chemical vapor deposition (MOCVD) reactor. Mo(CO)$_6$ (99.99 % purity, Sigma-Aldrich) and Re$_2$(CO)$_{10}$ (99.99 % purity, Sigma-Aldrich) precursor powders are loaded into stainless steel bubblers. The pressure inside the bubblers is held at 735 Torr while the temperature is maintained at 24 °C and 35 °C for Mo(CO)$_6$ and Re$_2$(CO)$_{10}$, respectively. During growth, mass flow controllers (MFCs) modulate the flow of high purity H$_2$ through the bubblers. By regulating the vapor pressure inside the precursor bubblers and the carrier gas flow, precursor delivery can be tuned below $10^{-5}$ sccm. Sulfur is directly injected into the reactor by flowing high purity H$_2$S (99.5%, Sigma-Aldrich) through a mass flow controller. C-plane sapphire (Cryscore Optoelectronic Ltd, 99.996 % purity) are cleaned by sonicating them in acetone, isopropyl alcohol, and DI water for 10 mins each. The substrates are then immersed in a commercial Piranha solution bath (Nanostrip, KMG Electronic Chemicals) for 20 mins at 80 °C. Substrates are subsequently rinsed in DI water and dried using N$_2$. Substrates are loaded in the hot zone of the reactor and allowed to soak at base pressure ($\approx 10^{-3}$ Torr) at 300°C for 15 mins to drive off any moisture prior to growth. The furnace is then filled with high purity Argon gas to the growth pressure of 50 Torr. After the chamber pressure has equilibrated the temperature is ramped up to 900 °C (the growth temperature) at a rate of 50 °C/min. 2 s.l.m. Ar is continuously flown through the chamber during growth to push precursor down the tube and deliver them over the substrate surface. A three-step (nucleation, ripening, and lateral growth) process is used to regulate nucleation rate and achieve smooth ML epitaxial films on sapphire. During the nucleation step $1.5 \cdot 10^{-3}$ sccm Mo(CO)$_6$ and 40 sccm H$_2$S are flown for 2 minutes. The nuclei created are then ripened into MoS$_2$ domains by ceasing Mo(CO)$_6$ delivery and maintaining a steady H$_2$S supply for 10 minutes. The MoS$_2$ domains are then grown laterally by reintroducing Mo(CO)$_6$ at half the flow rate of the nucleation step while still injecting 40 sccm H$_2$S in the reactor. Complete MoS$_2$ monolayer coalescence on c-plane sapphire is achieved after 24 minutes of lateral growth time.

**Atomic Force Microscopy (AFM):** Atomic force microscopy (AFM) is were performed using Bruker Dimension Icon instrument equipped with a ScanAsyst-Air (k = 0.4 N/m) tip in tapping mode. During high-resolution scans the tip scan rate is set at 0.5 Hz while maintaining a peak force setpoint of 1.00 nN and a peak force frequency of 2 kHz.

**X-ray Photoelectron Spectroscopy:** Physical Electronics Versa Probe II tool was used to obtain XPS spectra. A monochromatic Al Kα X-ray source (hv = 1486.7 eV) with a 200 µm spot size was used to interrogate the samples at high vacuum (<10$^{-6}$ Torr). High-resolution spectra are obtained at a pass energy of 29.35 eV and 0.125 eV energy step. An ion gun and floating electron neutralizer were used for charge neutrality and all component spectra were charge corrected to C1s spectrum at 284.8 eV. VBM were estimated by setting the Mo 3d$_{5/2}$ peak to 229.0 eV.

**TEM sample preparation:** A PMMA-mediated, wet-transfer method similar to the one described in the BGFET fabrication section below was used to transfer the MoS$_2$ films from the sapphire growth substrate to a 3 mm diameter Cu Quantifoil TEM grid. Rather than fishing out the PMMA/MoS$_2$ stack, a Büchner funnel filled with DI water is used to let the PMMA/MoS$_2$ film slowly drape onto the TEM grid to minimize film damage.

**High resolution scanning transmission electron microscopy (HR-STEM):** High resolution scanning transmission electron microscopy (STEM) was performed at 80 kV on a dual spherical aberration-corrected FEI Titan G2 60-300 S/TEM. All the STEM images were collected by using a high angle annular dark field (HAADF) detector with a collection angle of 42-244 mrad, camera length of 115 mm, beam current of 40 pA, beam convergence of 30 mrad.

**Raman Spectroscopy:** Raman spectra were measured in a Horiba LabRam HR Evolution VIS-NIR Raman system using excitation wavelengths of 633nm in ambient conditions.

**Absorbance Measurement:** The absorbance spectrum of undoped and doped ML MoS$_2$ directly grown on transparent sapphire is measured in Agilent/Cary 7000 spectrophotometer.

**STM, STS:** Undoped MoS$_2$ and Re-doped (0.1-5 at.%) MoS$_2$ samples were grown on QFEG on SiC. Subsequently the samples were transported through air and annealed at 250-300°C in ultra-high vacuum (≈2·10$^{-10}$ mbar). The STM/STS measurements were performed with a commercial low-temperature STM from Scienta Omicron operated at 5 K. STM topographic measurements were taken in constant current mode with the bias voltage applied to the sample. STS measurements were recorded using a lock-in amplifier HF2LI from Zurich Instruments running at 860 Hz. The tungsten tip was prepared on a clean Au(111) surface and confirmed to be metallic.

**PL and TRPL Characterization**: Steady-state PL measurements were collected in vacuum ($\approx 10^{-6}$ torr) using a microscope built around a closed-cycle helium cryostat (Montana Instruments, s100). The microscope used a 100 x 0.75 NA objective to focus the output of a continuous-wave laser (Oxxius LBX-445, $\lambda_{ex}$ = 445 nm: 10 µW) onto the sample. Following optical excitation, PL was collected by the same objective and separated from stray laser light using a 550 nm dichroic mirror (Thorlabs) and a 600 nm long-pass filter (Thorlabs). The PL was then coupled into an optical fiber and focused onto the slits of a spectrograph (Princeton Instruments HRS-300SS, grating 300 grooves/mm) before being detected by a back-illuminated CCD (Princeton Instruments, PIXIS-400BR). For temperature-dependent measurements, the sample was cooled from 300-77 K using the s100 cryostat.

For time-resolved PL (TRPL) measurements, a 1 MHz diode-pumped Yb:KGW laser (Carbide, Light Conversion Ltd, Vilnius, Lithuania) was used to pump an optical parametric amplifier (OPA) (Orpheus-F, Light Conversion) with $\approx$200 fs, 1030 nm pulses. 890 nm pulses produced by the OPA were focused into a beta barium borate (BBO) crystal to generate 445 nm laser pulses used to photoexcite the sample. The laser energy was 5 pJ/pulse for all time-resolved measurements. A 40 x 0.75 NA objective was used to focus the laser onto the sample and collect the sample's photoluminescence after excitation. The collected PL was dispersed into a spectrograph and detected using a streak camera (Hamamatsu, C14831-130). The temporal resolution of the instrument was $\approx$150 ps.

**First-Principles DFT Calculations:** First-principles calculations are performed using CP2K[61]. The GGA-PBEsol functional was used with GTH pseudopotentials[62]. DZVP basis set is used with a 900 Ha max. cut-off energy for the real-space integration. We construct a 10x10 supercell of single-layer hexagonal $MoS_2$ and substitute one Mo atom with a Re one (1 % defect concentration). The unit cell and internal degrees of freedom of the neutral systems are first relaxed; then we used these relaxed structures for an additional relaxation of the internal degrees of freedom of the charged and neutral systems this time with a 2D Poisson solver to avoid the spurious interactions between the layers out of plane. A homogeneous background charge of opposite sign is introduced to counterbalance the introduced charged state. Spin-polarization is included but is found to lead to negligible modifications of the densities of states.

The formation energy of the sulfur vacancies ($E_{form}$)

$$E_{form} = E_{V_S} + \mu_S - E_{MoS_2}$$

is computed, with $E_{form}$ is the formation energy of the S vacancy, $E_{V_S}$ and $E_{MoS_2}$ the (ground-state) total energies of MoS$_2$ with and without a V$_S$, respectively, and $\mu_S$ is the chemical potential of sulphur, taken here as half of the S$_2$ molecule energy. For the edge calculations, we construct a 9x8 supercell of MoS$_2$ and place it in a box with 10 Å of vacuum in this direction to isolate the edge and 16 Å of vacuum orthogonal to the 2D plane. We cut along the zigzag direction of ML MoS$_2$ and consider S-terminated edges. A V$_S$ is then introduced at the edge, while a Mo atom is substituted by a Re one (≈1.4 at.% doping concentration) in the closest vicinity of the V$_S$. Then, this substitution is done further away from the edge up to 11 Å deep into the model. During relaxation of the internal degrees of freedom (cell is fixed), the position of the atoms on the opposite edge are kept fixed to mimic the bulk of MoS$_2$ and to avoid spurious interactions between periodic images.

**Fabrication of MoS$_2$ BGFETs:** Following the growth of large area MoS$_2$, each as-grown film is transferred from the growth substrate to a 50 nm Al$_2$O$_3$/Pt/TiN/p++ Si substrate using the PMMA assisted wet transfer process. Initially, each sample was spin coated with PMMA and left in the desiccator overnight. Next, the edges of the samples were scratched using a razor blade. The samples were then placed in 1M NaOH at 90 °C to delaminate the film from the growth substrate. The films are rinsed in 3 water baths for 10 min each to remove any residual NaOH and then fished out with the target device substrate. The samples are then annealed at 50 °C and 70 °C for 10 min each to improve the adhesion of the film to the target substrate. The PMMA is stripped from the films by placing the samples in a 10 min acetone bath and are subsequently cleaned in a 10 min IPA bath. Channels are defined using e-beam lithography and isolated via dry reactive ion etching using SF6 at 5 °C for 30 s. After stripping the resist, source and drain terminals are defined using e-beam lithography. E-beam evaporation was implemented to deposit 40 nm nickel and 30 nm gold. Liftoff of the excess metal was done by placing the samples in an acetone bath at 50 °C for 30 mins and in an IPA bath for 15 mins.

**Electrical Characterization:** Characterization of the fabricated devices was carried out in a Cascade SUMMIT200 automated probe station with a Keysight B1500A parameter analyzer in atmospheric conditions. This probe station was primarily implemented to obtain device statistics for each film. Vacuum measurements were carried out in a Lake Shore CRX-VF probe station using the same parameter analyzer as mentioned above.

## Acknowledgements

R.T., Y.-C. L, and J.A.R. acknowledge funding from NEWLIMITS, a center in nCORE as part of the Semiconductor Research Corporation (SRC) program sponsored by NIST through award number 70NANB17H041. This material is based upon work supported by the National Science Foundation Graduate Research Fellowship Program under Grant No. DGE1255832. Any opinions, findings, and conclusions or recommendations expressed in this material are those of the author(s) and do not necessarily reflect the views of the National Science Foundation. L.H. and B.S. acknowledge funding from the European Research Council (ERC) under the European Union's Horizon 2020 research and innovation program (Grant agreement No. 948243). K.T.M., J.A.R., and J.B.A. acknowledge funding from the U.S. National Science Foundation Major Research Instrumentation program for development of the steady-state and time-resolved PL microscope in vacuum through award number DMR-1826790. E.M., B.V.T. and G.P. thank the Imec Industrial Affiliation Program (IIAP) for funding.

# Supporting Information
## Dilute Rhenium Doping and its Impact on Intrinsic Defects in MoS$_2$

Riccardo Torsi,[1] Kyle T. Munson,[2] Rahul Pendurthi,[3] Esteban A. Marques,[4,5] Benoit Van Troeye,[4] Lysander Huberich,[6] Bruno Schuler,[6] Maxwell A. Feidler,[1] Ke Wang,[7] Geoffrey Pourtois,[4] Saptarshi Das,[3] John B. Asbury,[2,†] Yu-Chuan Lin,[1,#] Joshua A. Robinson[1,2,7,8*]

1. Department of Materials Science and Engineering, The Pennsylvania State University, University Park, Pennsylvania 16802, United States
2. Department of Chemistry, The Pennsylvania State University, University Park, Pennsylvania 16802, United States
3. Department of Engineering Science and Mechanics, The Pennsylvania State University, University Park, Pennsylvania 16802, United States
4. Imec, Leuven 3001, Belgium
5. Department of Molecular Design and Synthesis, KU Leuven, Celestijnenlaan 200f - Postbox 2404, 3001 Leuven, Belgium
6. nanotech@surfaces Laboratory, Empa-Swiss Federal Laboratories for Materials Science and Technology, Dübendorf 8600, Switzerland
7. Materials Research Institute, The Pennsylvania State University, University Park, PA, 16802, USA
8. Department of Physics, The Pennsylvania State University, University Park, Pennsylvania 16802, United States

\# yul194@psu.edu

† jba11@psu.edu

\* jar403@psu.edu

**DFT model to determine Re dopant incorporation in MoS₂ grains**

To determine the favorable position of the Re dopant atom inside a grain, we rely on the computation of the Gibbs free energy:

$$\Delta G_{rxn} = G_{Re,center}(T) - G_{Re,edge}(T),$$

Where $G_{Re,center}$ and $G_{Re,edge}$ are the Gibbs free energies of the grains with one Re dopant atom at the center or at the edge, respectively. In practice, it is estimated based on a second-order expansion of the energy, i.e. on the total energy in the Born-Oppenheimer approximation and on the free energy of phonons.

To model grains utilize triangular nanoparticle of MoS₂ with 36 Mo atoms and 90 S atoms placed inside a box with 12Å vacuum out-of-plane and 10Åx10Å of vacuum in-plane. All edges of the nanoparticles are zigzag and S₂ terminated. The Re dopant is introduced as substitution of the Mo atom either at one of the 3 equivalent sites at the center of the nanoparticle (center configuration above) or at the one of the two equivalent sites at the middle of the edge (edge configuration). All the internal degrees of freedom are relaxed, and the final structures used for the subsequent phonon computations.

We then calculate the phonon frequencies within the harmonic approximation through the use of (second-order centered) finite difference for the interatomic force constants with a displacement of 0.005. Phonopy[1] is then used for the imposition of the acoustic sum rule and the estimation of thermodynamic quantities. Energy and force evaluations (as well as relaxation) are performed with CP2K[2] using a DZVP basis set and with a 500 Ha max. cut-off energy for the real-space integration.

**PL Peak fitting to determine electron density in rhenium doped MoS₂ films**

The photoluminescence of MoS₂ depends sensitively on the electron density within the material. We estimated the electron density of rhenium doped MoS₂ films using a model developed by Matsuda.[3] In brief, this model describes the populations of excitons and trions under steady-state conditions as

$$N_{ex} = \frac{G}{\Gamma_{ex}+k_{tr}} \qquad \text{(Eqn S1)}$$

$$N_{tr} = \frac{k_{tr}}{\Gamma_{tr}} \cdot \frac{G}{\Gamma_{ex}+k_{tr}} \qquad \text{(Eqn S2)}$$

where $N_{ex}$ is the neutral exciton population, $N_{tr}$ is the trion population, G is the optical generation rate for excitons, $\Gamma_{ex}$ and $\Gamma_{tr}$ are the decay rate constants for neutral excitons and trions, and $k_{tr}$ is the trion formation rate constant. For our analysis, $\Gamma_{ex} = 0.002\ ps^{-1}$, $\Gamma_{tr} = 0.02\ ps^{-1}$, and $k_{tr} = 0.5\ ps^{-1}$ based on prior transient absorption measurements.[4]

The PL intensity of excitons ($I_{ex}$) and trions ($I_{tr}$) is related to the exciton and trion populations according to

$$I_{ex} = \frac{AG\gamma_{ex}}{\Gamma_{ex}+k_{tr}} \qquad \text{(Eqn S3)}$$

$$I_{tr} = \frac{k_{tr}}{\Gamma_{tr}} \cdot \frac{AG\gamma_{tr}}{\Gamma_{ex}+k_{tr}} \qquad \text{(Eqn S4)}$$

where $\gamma_{ex}$ and $\gamma_{tr}$ are the radiative decay rate constants for excitons and trions and A is the PL collection efficiency. When $k_{tr} \gg \Gamma_{ex}$, the PL intensity of excitons ($I_{ex}$) and trions ($I_{tr}$) can be approximated as

$$I_{ex} \approx \frac{AG\gamma_{ex}}{k_{tr}} \qquad \text{(Eqn S5)}$$

$$I_{tr} \approx \frac{AG\gamma_{tr}}{\Gamma_{tr}} \qquad \text{(Eqn S6)}$$

A mass action model describes the relationship between neutral excitons, trions, and excess electrons as

$$\frac{N_{ex}n_{el}}{N_{tr}} = \frac{4m_{ex}m_{el}}{\pi\hbar^2 m_{tr}} k_b T \cdot \exp\left(-\frac{E_b}{k_b T}\right) \qquad \text{(Eqn S7)}$$

where, $n_{el}$ is the doped electron density, T is temperature, $k_b$ is the Boltzmann constant, $E_b$ is the trion binding energy (~40 meV), and $m_{ex}$, $m_{tr}$, and $m_{el}$ are the effective masses of excitons (0.8 $m_0$), trions (1.15 $m_0$), and electrons (0.35 $m_0$), respectively.[3] Here $m_0$ is the free electron mass.

From Eqn. S7, the doped electron density of a MoS$_2$ film is estimated from the spectral weight of trion PL according to

$$\frac{I_{tr}}{I_{total}} = \frac{\frac{\gamma_{tr} N_{tr}}{\gamma_{ex} N_{ex}}}{1 + \frac{\gamma_{tr} N_{tr}}{\gamma_{ex} N_{ex}}} \approx \frac{4.4 \cdot 10^{-14} n_{el}}{1 + 4.4 \cdot 10^{-14} n_{el}} \qquad \text{(Eqn S8)}$$

where $I_{X^-}/I_{total}$ is the trion PL spectral weight. $\gamma_{tr}/\gamma_{ex}$ was calculated to be ~0.1 from the intensity ratio of the trion and A-exciton PL curves.

Figure S9 shows the relationship between trion PL intensity and electron density obtained from Eqn. S8. Figure S9 also displays $I_{X^-}/I_{total}$ values calculated by fitting the PL spectra in Figure 4A. The comparison demonstrates that Re atoms act as n-type dopants that increase the electron concentration within the material.

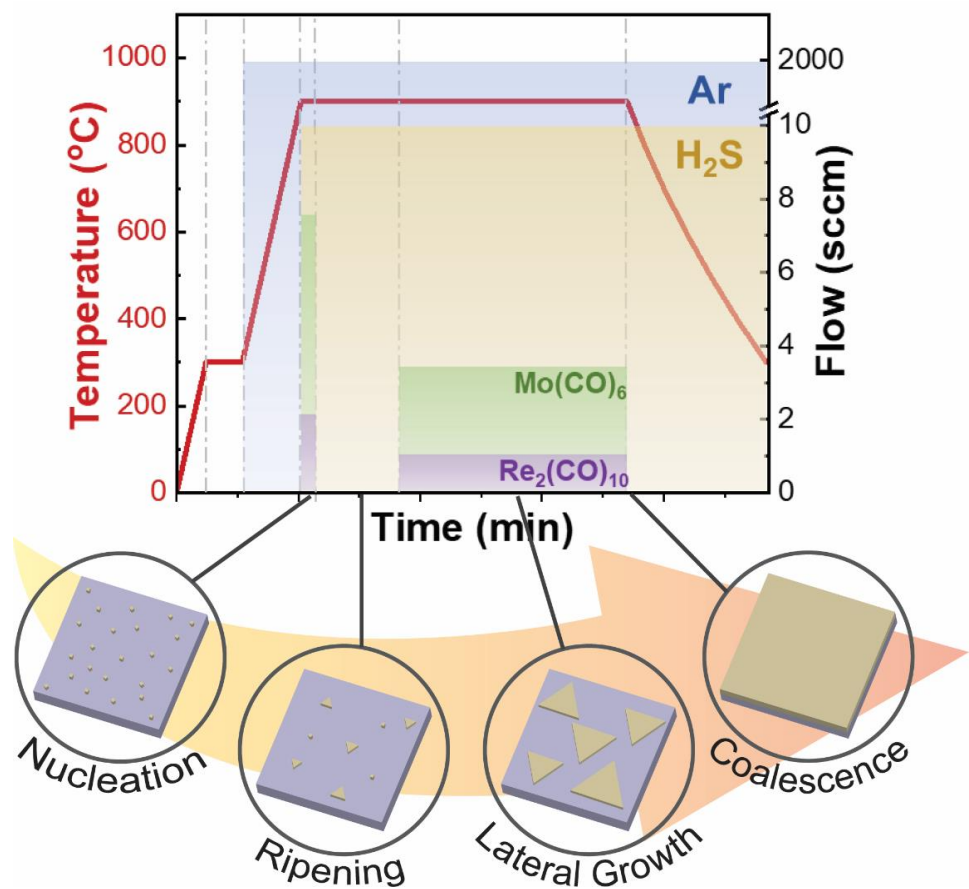

**Figure S1.** Illustration of the multi-step MOCVD growth method adopted in this study for synthesizing ML Re-MoS$_2$ films. By separating the growth into nucleation, ripening, and lateral growth stages, nucleation density and growth rate can be tightly regulated to achieve smooth monolayer films with minimal bilayer coverage.[5]

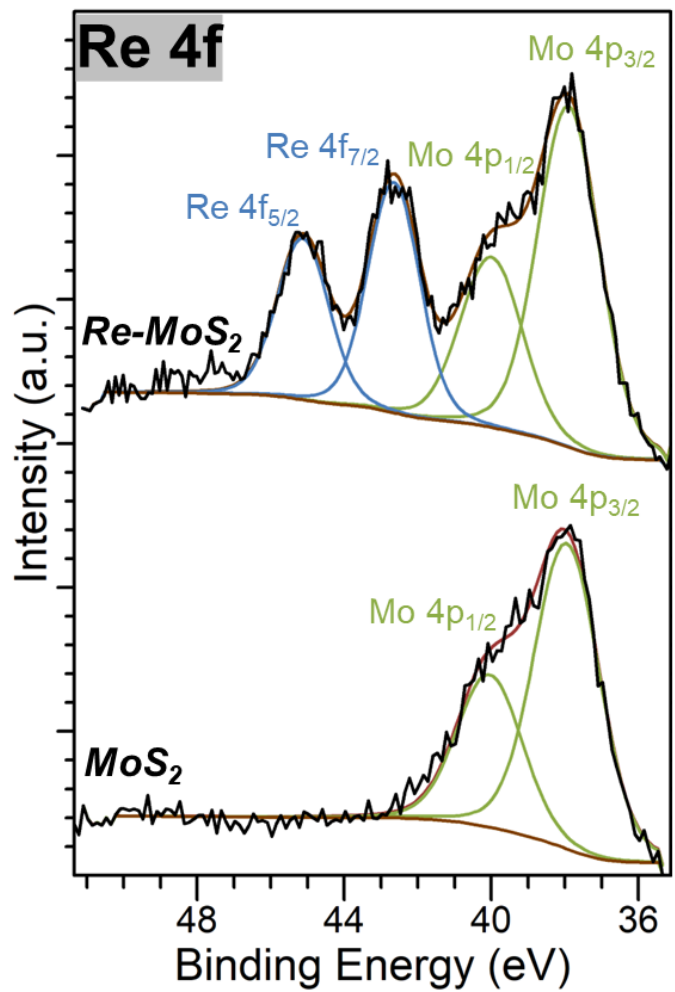

**Figure S2.** The high-resolution XPS Re binding energy regime of pristine ML MoS$_2$ and ML Re-MoS$_2$ with 6 at.% Re. Fitting of the Re *4f* peaks shown in blue are used to estimate Re content in the films.

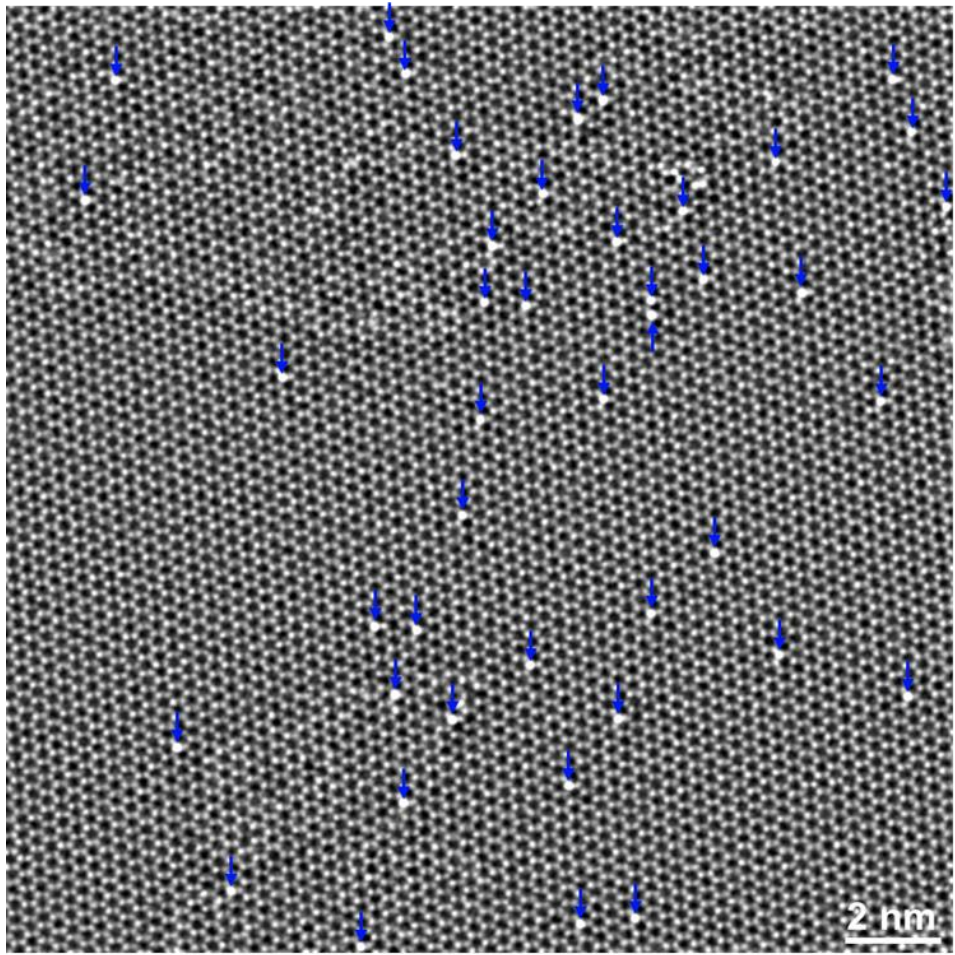

**Figure S3.** High-resolution Z-contrast STEM image of ML Re-MoS$_2$. 43 Re atoms were counted over a 20 x 20 nm$^2$ area, which yields a Re content of ~ 1.1 at.%. The calculated compositional value is consistent with that extracted from the XPS analysis for the same sample (~1.3 at.%).

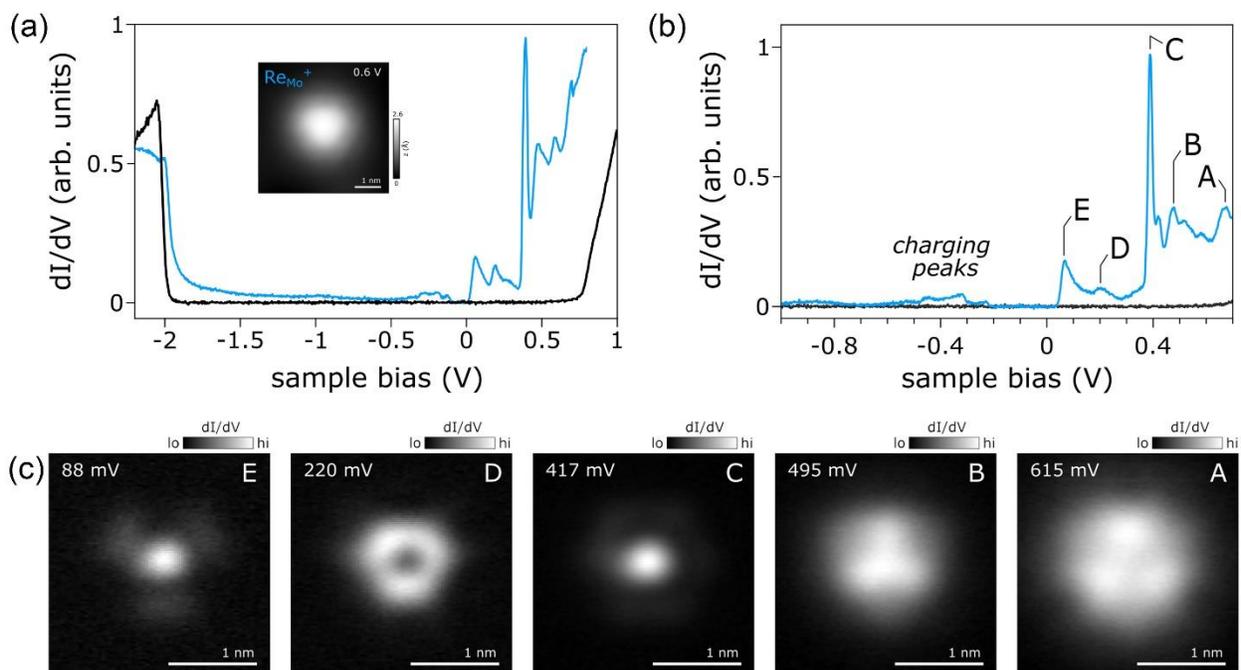

**Figure S4.** Electronic structure of Re in MoS$_2$ revealed by STS measurement at 5 K: (a,b) Differential conductance (dI/dV) spectroscopy (V$_{mod}$ = 10 mV) recorded on bare monolayer MoS$_2$ (black) and in the center of a single Re$_{Mo}^+$ dopant (blue). Inset: STM topography (V = 0.6 V, I = 100 pA) of Re$_{Mo}^+$. (c) *dI/dV* maps (V$_{mod}$ = 20 mV) of the main defect resonances labelled in (b).

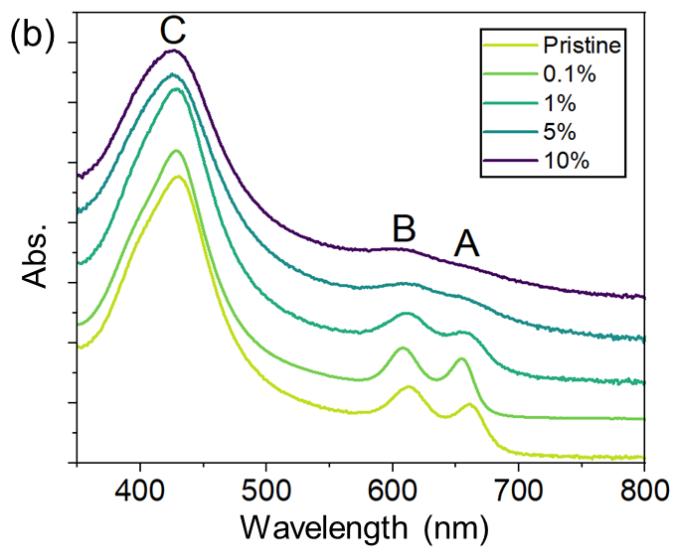

**Figure S5.** Optical absorbance of ML Re-MoS$_2$ as a function of Re concentration. (a) Raman spectral evolution of the LA, $E'$ and $A'_1$ mode of ML MoS$_2$ with 0, 0.1, 1, 5, and 10 at.% Re under 532 nm excitation wavelength. (b) Absorbance spectrum of ML Re-MoS$_2$ with increasing Re content.

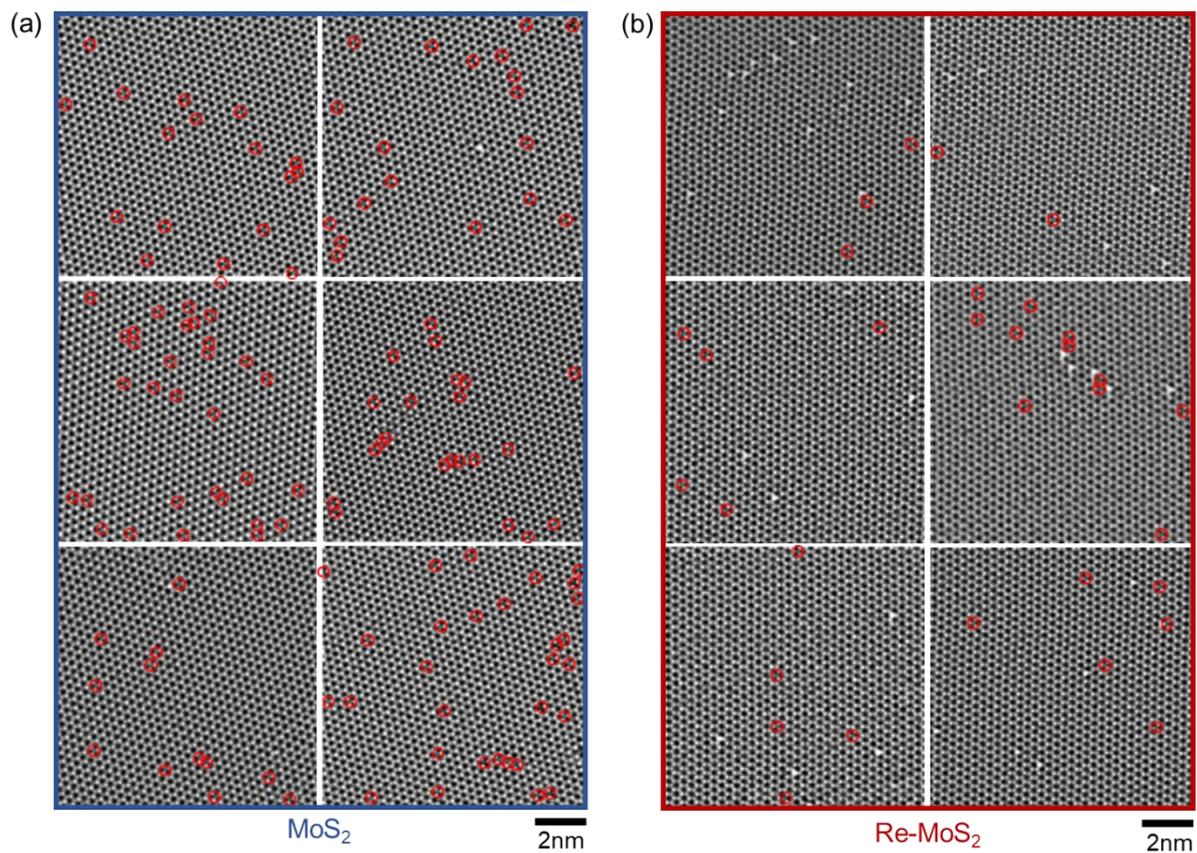

**Figure S6.** Defect density analysis for six 100 nm² Z-STEM images from both (a) undoped MoS$_2$ and (b) Re-doped MoS$_2$. Localized weak Z-contrast intensity at the sulfur sites of monolayers caused by vacancies and impurities are marked with red circles.

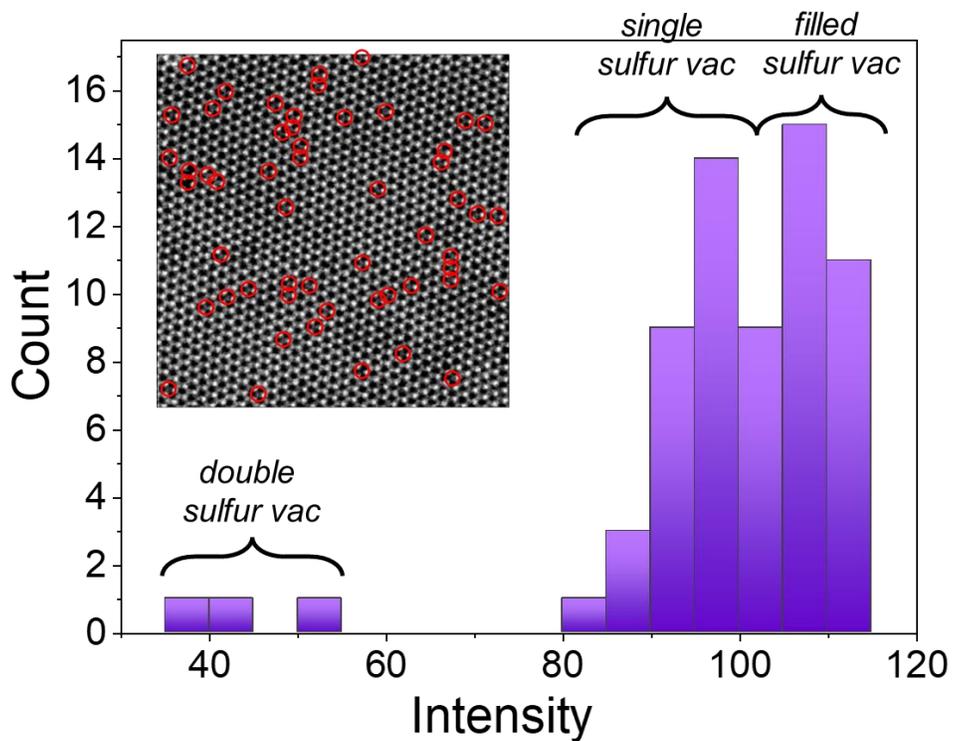

**Figure S7.** Histogram of Z-STEM intensity for S-site defects in pristine MoS$_2$ showing the presence of single sulfur vacancies, filled vacancies, and a few double sulfur vacancies.[6]

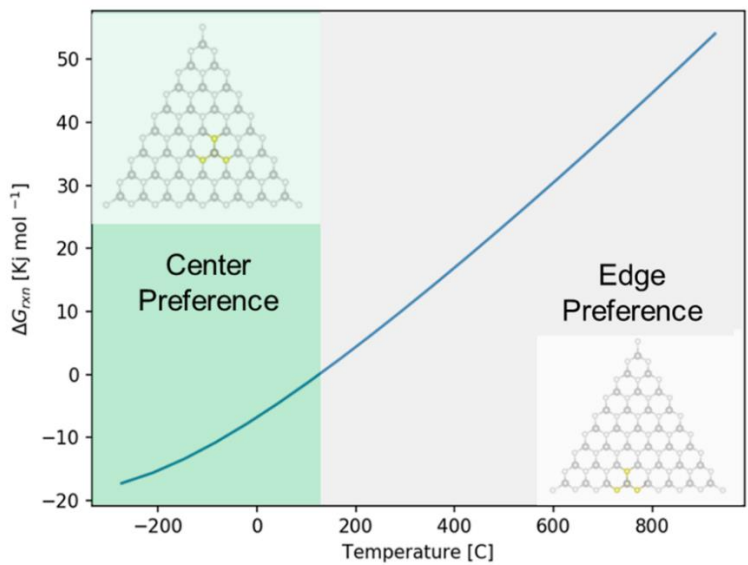

**Figure S8.** Formation energy for Re-Mo substitution in $MoS_2$ as a function of temperature showing regions for center and edge preference as calculated through DFT calculations. The plot clearly points to the tendency of Re dopant atoms to attach to the growth front of $MoS_2$ grains.

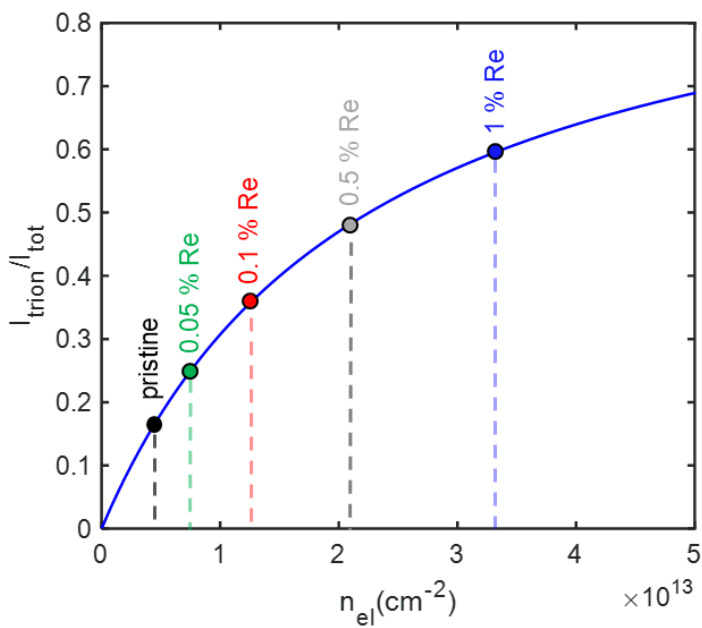

**Figure S9.** Trion PL spectral weight ($I_{X^-}/I_{total}$) plotted as a function of electron density ($n_{el}$) obtained from Eqn. S8. $I_{X^-}/I_{total}$ values calculated by fitting the PL spectra of 0 (pristine), 0.05, 0.1, 0.5, and 1 at.% Re-$MoS_2$ films.

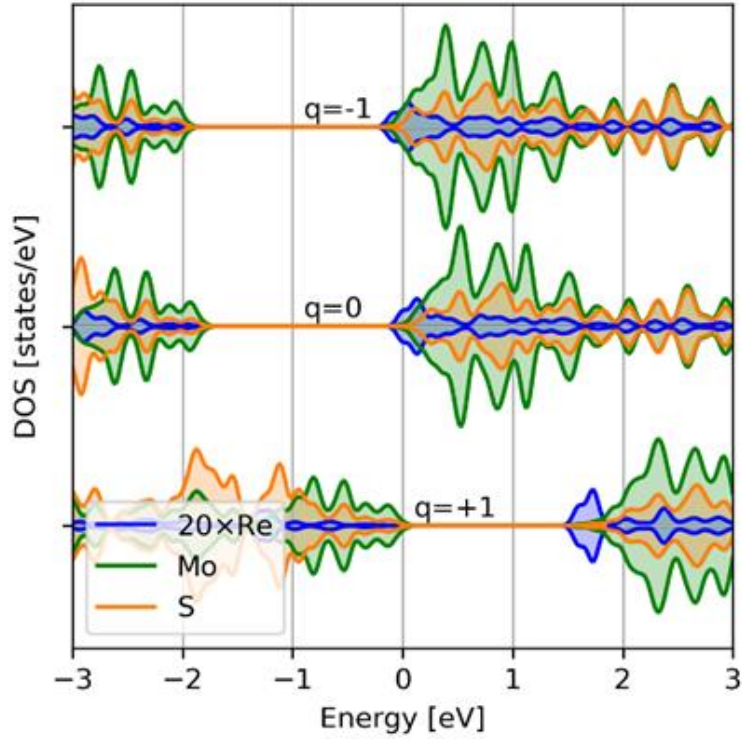

**Figure S10.** Projected density of states for defect-free Re-MoS$_2$ in different charge states.

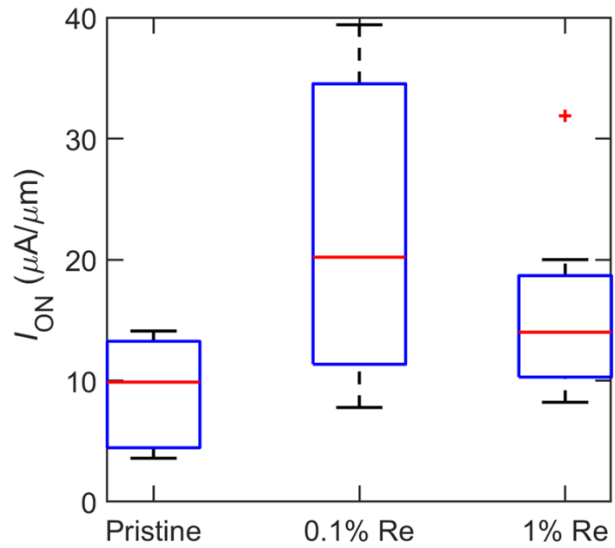

**Figure S11.** Extracted $I_{ON}$ at $V_{DS}$ = 5 V and $V_{BG}$ = 10 V for ten BGFET devices on 0 (pristine), 0.1, and 1.0 at.% Re-MoS$_2$.

**Table S1.** Fitting of time-resolved photoluminescence decay curves for 0 (pristine), 0.1, and 1.0 at.% Re-MoS$_2$ films using the bi-exponential equation (S3). The exciton and defect-related recombination contributions can be de-convoluted highlighting a ≈10x reduction in the long-lived, defect-mediated recombination pathway ($\tau_2$) when Re is incorporated in the MoS$_2$ lattice.

$$D(t) = ae^{-\frac{t}{\tau_1}} + (1-a)e^{-\frac{t}{\tau_2}} \quad (S3)$$

| Sample | a | $\tau_1^{Exciton}$ (ps) | $\tau_2^{Defect}$ (ps) | $t_{ave}$ (ps) |
|---|---|---|---|---|
| Pristine | 0.89 | 180 | 1330 | 307 |
| 0.1 at.% Re | 0.84 | 113 | 166 | 121 |
| 1.0 at.% Re | 0.84 | 80 | 150 | 91 |

**Table S2.** Key electrical properties of the MoS$_2$ BGFETs for the various Re doping concentrations extracted at $V_{DS}$ = 5 V. The table highlights moderate enhancement in both average $I_{ON}$ and mobility values with Re doping. The improved transport properties can be attributed to a reduction in the S-site defect density with Re- incorporation as no discernable n-type doping is discerned from $V_{TH}$ values.

| | | $V_{TH}$ (V) | $I_{ON}$ (μA/μm) | $I_{ON}/I_{OFF}$ | $\mu_e$ (cm$^2$/V-s) | $SS_e$ (mV/dec) |
|---|---|---|---|---|---|---|
| Pristine | Average | -2.19 | 7.34 | 5.70E+07 | 4.08 | 694 |
| | Std. Dev. | 0.89 | 5.03 | 6.86E+07 | 1.67 | 309 |
| 0.1 at.% Re | Average | -2.46 | 19.6 | 7.80E+07 | 8.74 | 788 |
| | Std. Dev. | 0.64 | 12.7 | 5.50E+07 | 3.52 | 148 |
| 1.0 at.% Re | Average | -1.95 | 15.8 | 4.10E+07 | 6.97 | 635 |
| | Std. Dev. | 0.5 | 8.05 | 2.79E+07 | 2.81 | 121 |